\documentclass{iopart}
\usepackage{graphicx}
\usepackage{amssymb}
\usepackage{hyperref}

\newcommand\eqref[1]{\eref{#1}}

\begin{document}

\title{Information temperature as a parameter of random sequence complexity} 

\author{ O.~V.~Usatenko}
\address{A. Ya. Usikov Institute for Radiophysics and
Electronics Ukrainian Academy of Science, 12 Proskura Street, 61805
Kharkiv, Ukraine\\
Department of Physics, University of Florida, Gainesville, FL 32611-8440, USA}
\ead{usatenkoleg@gmail.com}

\author{ G.~M.~Pritula}
\address{A. Ya. Usikov Institute for Radiophysics and
Electronics Ukrainian Academy of Science, 12 Proskura Street, 61805
Kharkiv, Ukraine}
\ead{pritula.galina@gmail.com}

\begin{abstract}

In this study, we continue our exploration of the concept of information temperature as a characteristic of random sequences.  We describe methods  for introducing the information temperature in the context of binary high-order Markov chain with step-wise memory, and investigate the application of the temperature as a parameter of the sequence complexity. We aim to define complexity based on the derivative of entropy with respect to information temperature, drawing an analogy to thermodynamic heat capacity. The maximum complexity of a random sequence is achieved when its information "heat capacity" approaches its highest possible value, which is directly influenced by the sequence memory depth . We also discuss the potential of utilizing information temperature as an indicator of the intellectual level  exhibited by any text-generating agent.

\end{abstract}

\vspace{2pc}

%%%%%%%%%%%%%%%%%%%%%%%%%%%%%%%%%%%%%%%%%%%%
\section{Introduction}
%%%%%%%%%%%%%%%%%%%%%%%%%%%%%%%%%%%%%%%%%%%

The description and understanding of the complexity of natural and artificial dynamical systems remains an open problem in science~\cite{Bennett,Thurner,Ladyman,Rong,Lavazza,Illiashenko}.
In connection with the rapid development of neural networks and artificial intelligence, the need to evaluate progress in the field of AI arises, including the possibility of comparing various models and devices that simulate intelligence with each other and with the real  human intelligence.

Quantitative evaluation of artificial intelligence faces currently several important challenges. A fundamental problem in artificial intelligence is that nobody really knows what intelligence is. There is no standard definition of what exactly constitutes intelligence. We do not have unified models of an artificially intelligent system. It is believed that intelligence encompasses a range of aptitudes, skills, and talents. Despite all these complications, artificial intelligence systems are improving, and currently well-known examples include self-driving cars, image classifiers, translators, chess and go game simulations, different content generators, including ChatGPT bot, among many others used in medicine, industry, and management.

Recourse to various existing definitions of human intelligence does not simplify much the problem of comparing intelligences. Human activity is so multifaceted and versatile that it is hardly possible to give a comprehensive and complete definition for human intelligence as well as to invent its universal measure. It is difficult to compare who is more gifted and brilliant writer or scientist, designer or artist, actor or director. Usually this raises the question how one can compare manifestations of intelligence in different fields of activity. One of the possible ways to do this is, for example, to \emph{consider intelligence as the capability to produce meaningful texts}.

In this work, we propose a new quantitative assessment for evaluating the intellectual capabilities of AI simulators and living beings based on the concept of information temperature. Specifically, we consider intelligence as the ability to produce coherent and meaningful text, and we propose a measure that seeks to capture this ability.   To assess the complexity of written language, we utilize the information temperature and introduce the information "heat capacity", which is the derivative of entropy with respect to this temperature.

The idea of the possibility and usefulness of introducing the concept of information temperature for texts considered as random correlated sequences was first proposed by Mandelbrot \cite{Mandel}.  Afterwards  the concept of ``text temperature'' was applied to linguistic analysis of the texts \cite{Campos,Kosmidis,Rego,Chang} under the assumption that human language could be described as a physical system within the framework of equilibrium statistical mechanics. This approach was used to analyze the high-frequency words with the use of the Boltzmann distribution~\cite{Miyazima} and, in a different way, to consider
the low-frequency vocabulary \cite{Rovenchak}. Our aim is to examine the information temperature as a complexity measure of random symbolic sequences represented by binary high-order Markov chains. 

The rest of the paper has the following structure. In Section \ref{GenDef} we provide a  brief description of the exactly solvable binary high-order Markov chain with step-wise memory and ways of introducing the information temperature. In Section~\ref{Entropy} we present  simple examples of the application of the concept of information temperature: the dependence of the entropy and complexity of the system on the information temperature as a parameter of ordering of the system. This Section also contains a discussion of the use of  information temperature to characterize information and intellectual complexity of both people and artificial intelligence systems. It assumes that these systems can be represented through written or generated texts. Section ~\ref{Conclusion} concludes the paper.

%%%%%%%%%%%%%%%%%%%%%%%%%%%%%%%%%%%%%%%%%%%%
\section{Information temperature}\label{GenDef}
%%%%%%%%%%%%%%%%%%%%%%%%%%%%%%%%%%%%%%%%%%%
In this section, we provide a brief introduction to the concept of information temperature. A more comprehensive explanation can be found in paper ~\cite{UsMPYa}.

In general, a possibility of introducing the information temperature can be explained by a coarse-grained description~\cite{Muntean} of the temporal evolution of dynamical system. By the coarse-grained description, we mean the division of the phase space of a dynamic system into cells and the subsequent representation of the true dynamics as a sequence of symbols representing the cells in which the point of the phase trajectory was located at different moment in time.

One of the methods for introducing information temperature, which we call the method of equivalent correspondences (EC), is based on the fact that every binary Markov chain is equivalent to some binary two-sided chain \cite{AMUYa} which, in its turn, can be viewed as the Ising sequence at a fixed temperature where the probability of a configuration is given by the Boltzmann distribution.

The second method makes use of  the traditional entropy based  thermodynamic definition of temperature with direct calculation of  the block entropy and energy of Markov chain in the pair-interaction  approximation. This method does not suppose a mapping of random sequence on the Boltzmann exponential distribution and describes a larger class of random sequences as compared to the previous EC method.

Both approaches give the same result for the case of nearest neighbor spin/symbol interaction, but the method of correspondence of Markov and Ising chains becomes very cumbersome for the chain orders $N>3$.
We present here  the first method of introducing information temperature for the simple case of ordinary Markov chain to provide the basics of concept. The rest of necessary details one can find in~\cite{UsMPYa}.

Let us consider an infinite random stationary ergodic sequence $\mathbb{S}$  of symbols-numbers $a_{i}$,
\begin{equation}
\label{RanSeq} \mathbb{S}= ..., a_{0}, a_{1},a_{2},...
\end{equation}
taken from the binary alphabet $a_{i}\in \mathcal{A}$:
\begin{equation}\label{alph}
 \mathcal{A}=\{0,1\},\,\, \,\, i \in
\mathbb{Z} = \{...,-1,0,1,2...\}.
\end{equation}
A higher number of states $\mathcal{A}$ can be handled by binary coding. We use the notation $a_i$ to indicate a position $i$ of the symbol $a$ in the chain and the unified notation $\alpha^k$ to stress the value of the symbol $a\in \mathcal{A}$.

We suppose that the symbolic sequence $\mathbb{S}$ is a \textit{high-order Markov chain}. The sequence $\mathbb{S}$ is a Markov chain if it possesses the following property: the probability of symbol~$a_i$ to have a certain value $\alpha^k \in \mathcal{A}$ under the condition that {\emph{all}} previous symbols are fixed depends only on $N$ previous symbols,
\begin{eqnarray}\label{def_mark}
 P(a_i=\alpha^k|\ldots,a_{i-2},a_{i-1})=
P(a_i=\alpha^k|a_{i-N},\ldots,a_{i-2},a_{i-1}).
\end{eqnarray}

For a \emph{two-sided random chain} the conditional probability that symbol~$a_i$ is equal to unity, under condition that the \emph{rest} of symbols in the chain are fixed, can be presented in the form~\cite{AMUYa},
\begin{equation} \label{2}
P(a_i=1|A_i^-,A_i^+)=\displaystyle\frac{P(a_i=1,A_i^+|A_i^-)}
{P(a_i=1,A_i^+|A_i^-)+P(a_i=0,A_i^+|A_i^-)},
\end{equation}
where
$A_i^-=(a_{i-N}\ldots,a_{i-2},a_{i-1})$ and $A_i^+=(a_{i+1},a_{i+2},\ldots,a_{i+N})$
are previous and next words of the length N with respect to symbol $a_i$.
Here the two-sided conditional probability $P(a_i=1|A_i^-,A_i^+)$ is expressed
by means of the Markov-like probability functions $P(a_i,A_i^+|A_i^-)$. 

For the stationary Markov chain, the probability $b(a_{1}a_{2}\dots
a_{N})$ of occurring a certain word $(a_{1},a_{2},\dots ,a_{N})$
satisfies the condition of compatibility for the Chapman-Kolmogorov
equation (see, for example, Ref.~\cite{gar}):

\begin{equation}
b(a_{1}\dots a_{N})=\sum_{a=0,1}b(aa_{1}\dots a_{N-1})P(a_{N}\mid a,a_{1},\dots
,a_{N-1}).  \label{10}
\end{equation}
In works~\cite{UYa} and~\cite{UYaKM}, we have introduced the model
Markov chain for which the conditional probability $p_{k}$ of
occurring the symbol ``0'' after the $N$-symbol words containing $k$
unities, e.g., after the word
$(\underbrace{11...1}_{k}\;\underbrace{00...0}_{N-k})$, is given by
the following expression:
\begin{equation}
p_{k}=P(a_{N+1}=0\mid \underbrace{11\dots
1}_{k}\underbrace{00...0}_{N-k})=\frac{1}{2}+\mu \left(1-\frac{2k}{N}\right).  \label{1}
\end{equation}
where $\mu$ is the model parameter, $|\mu|<1/2$.

For the ordinary one-step Markov chain with nearest-neighbor interaction using Eq.~\eqref{1} with $N=1$,  we easily obtain the conditional probabilities
\begin{eqnarray}\label{111}
&& P(a_i=1,a_{i+1}=1|a_{i-1}=1)=\left(\frac{1}{2}+\mu\right)^2,\\[6pt]
&&P(a_i=0,a_{i+1}=1|a_{i-1}=1)=\left(\frac{1}{2}-\mu\right)^2.\nonumber
\end{eqnarray}

Considering the equivalence of Markov chain to the two-sided random sequence, we apply
formula~\eqref{2} to the word $(a_{i+1},1,a_{i-1})$:
\begin{equation}\label{TwoOne}\hspace{-5mm}
P(a_i=1|a_{i-1},a_{i+1})=\frac{P(a_i=1,a_{i+1}|a_{i-1})}{P(a_i=1,a_{i+1}|a_{i-1})+
P(a_i=0,a_{i+1}|a_{i-1})}.
\end{equation}

The Boltzmann distribution for the corresponding Ising model is,
\begin{eqnarray}\label{P Ising}
P(a_i=1|a_{i-1}=1,a_{i+1}=1) =\frac{\exp(2\varepsilon/T)}{\exp(2\varepsilon/T)+\exp(-
2\varepsilon/T)}.
\end{eqnarray}

The binary variables $a_i=\{0,1\}$ of the Markov chain is related to the spin Ising's variables $s_i=\{-1,1\} = \{\downarrow,\uparrow\}$ by the equality
$s_i = 2 a_i - 1$, and the energy of the spins interaction is of the form: $\varepsilon_{\uparrow
\uparrow }=\varepsilon_{\downarrow\downarrow}=-\varepsilon$,
$\varepsilon_{\uparrow \downarrow
}=\varepsilon_{\downarrow\uparrow}=\varepsilon>0.$

Using Eqs.~\eqref{111} - \eqref{P Ising}, we have
\begin{equation}
  \mu = \frac{1}{2} \tanh {\left( \frac{1}{\tau} \right)}, \quad\quad
  \beta=\frac{1}{\tau} = \frac{1}{2} \ln \frac{1+2\mu}{1-2\mu}.
 \label{MuVsT}
\end{equation}
Here we have introduced the information temperature $\tau=
T/\varepsilon$ of the ordinary, one-step, Markov chain and $\beta=\varepsilon/T$
is the inverse info-temperature. So, for $\tau \rightarrow {\pm\infty}$  we have $\varepsilon/T \simeq 2\mu
\rightarrow 0$, and $\tau \rightarrow \pm 0$ when $\mu \rightarrow
\pm {1/2}$. The negative values of $\tau$
describes an anti-ferromagnetic ordering of spines or symbols\,
"0"\, and "1". We can say that $\tau$ is the temperature $T$
measured in unites of  energy $\varepsilon$, $\tau={T}/
\varepsilon$.

Considered here model of the Markov chain \eqref{1} describes the so-called non-biased sequences in which the number of symbols ``0'' and ``1''  is equal to each other and the parameter $\mu$ describes the strength of correlations. The results for this model can be quite easily generalized to the biased chains.
%
%%%%%%%%%%%%%%%%%%%%%%%%%%%%%%%%%%%%%%%%%%%%
\section{Information temperature as a parameter of entropy and complexity}\label{Entropy}

The above presented is a part of our previous finding ~\cite{UsMPYa} on the information temperature which should be considered as a macroscopic statistical characteristic of stationary ergodic random sequences. The temperature is one of the most important macroscopic parameters in statistical physics, and we believe that the concept of information temperature should have a deep meaning in relation to random sequences as well. In this section of the work, we present two simplest examples of the application of the concept of information temperature.

The relationship between entropy, complexity of a system, and its ordering parameter was earlier discussed by several authors ~\cite{Hogg,Li,Lopez}. However, this parameter lacked a clear and precise name or definition, being alternatively referred to as the "degree of disorder" of the system~\cite{Hogg}, the "order-disorder" parameter ~\cite{Li}, or the parameter ranging from an "ideal crystal to ideal gas" ~\cite{Lopez}. If earlier the question of which specific order parameter determines the entropy of a random symbolic system could be interpreted ambiguously, now we can provide a natural answer to this question: the entropy depends on the information temperature.
\subsection{Entropy as function of information temperature}\label{EntrTau}

\begin{figure}[h!]
\begin{centering}
\scalebox{0.7}[0.7]{\includegraphics{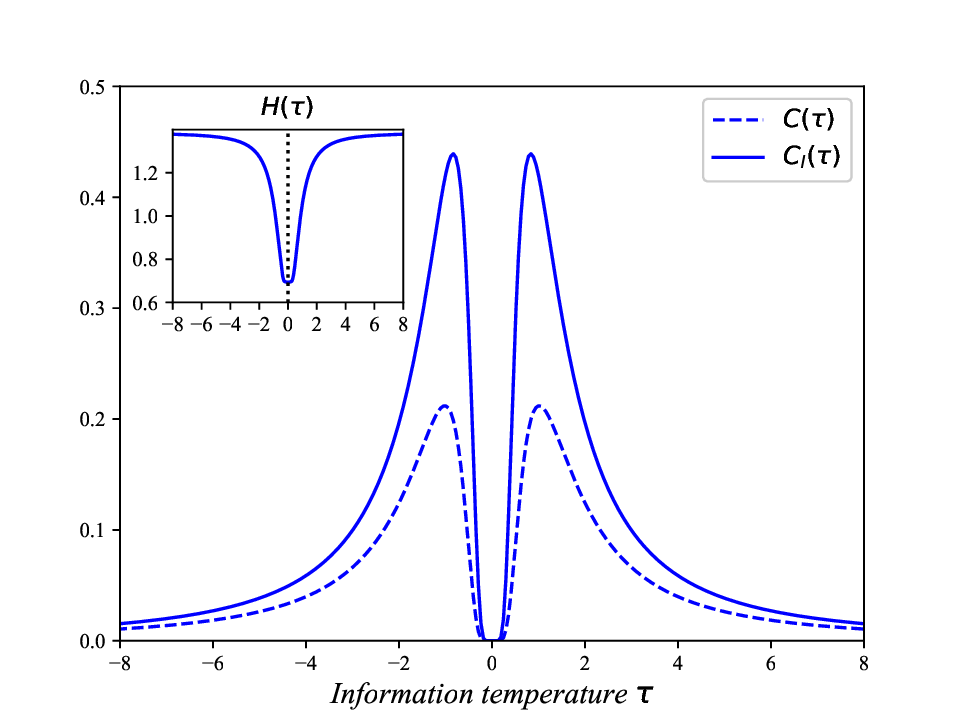}}
\caption{The dependence of complexity $C(\tau)$ and specific information heat capacity $C_I(\tau)$ on the information temperature $\tau$ for the 1-step Markov chain.} \label{Fig1}
\end{centering}
\end{figure}

The entropy per one bond $H_2$ for the ordinary one-step Markov chain can be easily found 
(see~\cite{UsMPYa}):
\begin{equation}
  H_2 = -2 P_{00}\ln P_{00}-2 P_{01}\ln P_{01}= \ln 2 \nonumber
  \end{equation}
\begin{equation}\label{S}
 -(1/2+\mu) \ln (1/2+\mu)-(1/2-\mu) \ln (1/2-\mu).
\end{equation}
Here $P_{ij}$ are the probabilities of the words (ij) of the length two  occurring:
\begin{equation}\label{00}
P_{00}=P_{11}=(1/2+\mu)/2,\,\quad P_{01}=P_{10}=(1/2-\mu)/2,
\end{equation}
and the parameters $\mu$ is defined by Eq.~\eqref{MuVsT}.
As a result, we get the entropy as a function of info-temperature,
\begin{eqnarray}\label{H2}
H_2=\ln 2 + \frac{\ln(1+\exp(-2\beta))}{1+\exp(-2\beta)}+ \frac{\ln(1+\exp(2\beta))}{1+\exp(2\beta)}.
\end{eqnarray}
This is a very natural and expected expression that approaches its minimum and maximum values as $\tau=1/\beta$ approaches zero or infinity, respectively. This result is presented in the insert of Fig.~\ref{Fig1}.

\subsection{Complexity as function of information temperature}\label{ComplTau}

Earlier it was pointed out~\cite{Hogg,Li} that the algorithmic (entropic, Kolmogorov) complexity~\cite{Kolmogorov} does not correspond to our intuitive understanding of complexity. In this regard, the paper~\cite{Lopez} proposes a definition that discriminates complexity in the region of high disorder or, in our terms, when $\tau \rightarrow \infty$.
The complexity of system with $N$ accessible states is introduced simply as
 the interplay between the information stored in the system and its disequilibrium.
In our case, this definition has the following form:
\begin{eqnarray}
C = h D, \quad\quad D=\sum_{ij ={0,1}} \left(P_{ij} -\frac{1}{4}\right).
\end{eqnarray}
Here $h =H_2-H_1$ is the Shannon entropy, or the entropy per symbol,
 $H_1=\ln 2$, $D$ is the so called disequilibrium factor~\cite{Lopez}, equal in our case to the value $D=\mu^2$, obtained with using Eq.~\eqref{00}. The final result for the complexity expressed via the information temperature $\tau$ is
\begin{eqnarray}\label{h D}
C(\tau)= \tanh^2(1/\tau)\left[ \frac{\ln(1+\exp(-2/\tau))}{1+\exp(-2/\tau)} + \frac{\ln(1+\exp(2/\tau))}{1+\exp(2/\tau)}\right].
\end{eqnarray}

This result is presented by the  dashed curve in Fig.~\ref{Fig1}.  In general, a properly defined complexity measure should reach its maximum at some intermediate level between the order of the completely regular and the disorder of the absolutely random~\cite{Majtey}. We see that the maximum of complexity $C(\tau)$ is reached at a finite value of the temperature, not at $T\rightarrow \infty$, as it takes place for $H$ in the insertion of Fig~\ref{Fig1}.

Another possibility to introduce the complexity measure which may be considered as more intrinsic quantity of random sequence and does not use an additional interplay between the information stored in the system and its disequilibrium, is
\begin{equation}
 C_{I}=\tau dh/d{\tau}.
 \end{equation}

In the statistical thermodynamics this quantity is known as a heat capacity, here we will call it the specific \emph{information heat} capacity or specific \emph{information heat} because the value $h$ is the excess of information entropy after one next symbol generation.

Both dependencies $C(\tau)$ and $C_{I}(\tau)$ for ordinary Markov chain are  presented  in Fig.~\ref{Fig1}.
At positive $\tau$, the function  $C_{I}(\tau)$ has the same \emph{one hump property} as the complexity $C(\tau)$ given by Eq.~\eqref{h D}.

\begin{figure}[h!]
\begin{centering}
\scalebox{0.7}[0.7]{\includegraphics{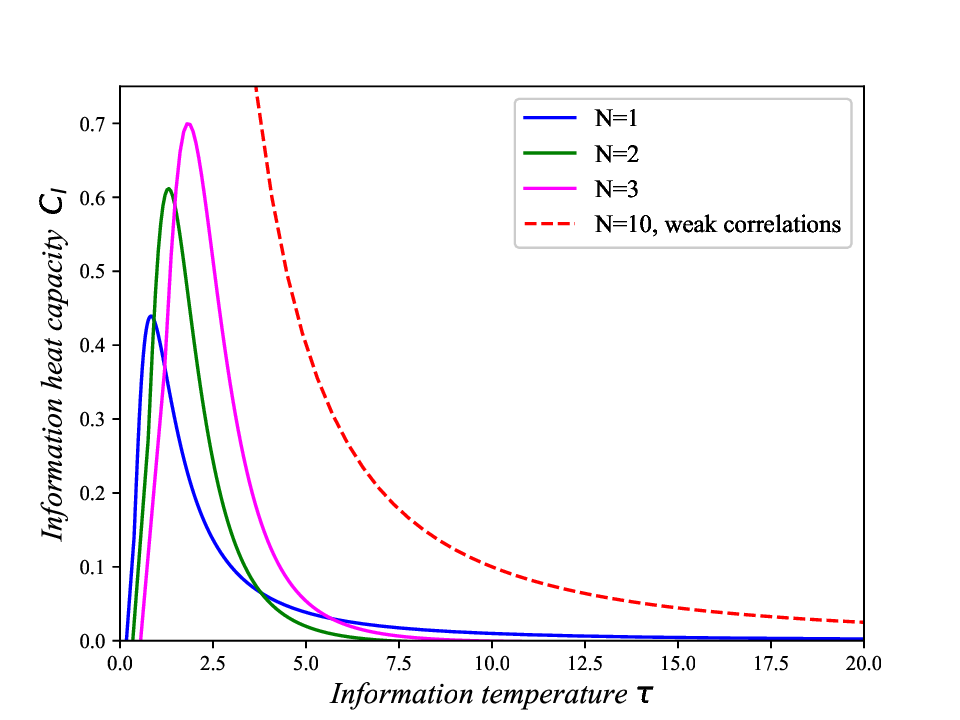}}
\caption{The specific information heat capacities $C_{I}(\tau)$ for step-wise Markov chains of orders $N=1,2,3$ and for large $N$ in the case of weak correlations.} \label{Fig2}
\end{centering}
\end{figure}

In the same way the specific information ``heat'' capacity can be obtained for the $2$ and $3$-step chains. In Fig.~\ref{Fig2}  we present  the plots of  $C_{I}(\tau)$ for the chains of order $N\le3$  and for the chain of the $10$-th order with weak correlations (note that for large $N$, we have used the entropy-based approach, see ~\cite{UsMPYa} for details).
A random sequence is most complex when its specific information heat capacity is close to the maximum, which value increases with increasing of the memory length.

Since the memory depths of any meaningful text usually reaches large values ($N \sim 10^2-10^5$, see, e.g.,~\cite{UYaKM}), the complexity of such texts should be presented by curves similar to the dashed one.

Thus, we can note that not only the information temperature but also the specific information heat can play an important role in the description of stationary random sequences which is a further advancement in the pursuit of understanding the equivalence of thermodynamic and information entropy.
This perspective appears to hold significant promise, particularly if one can grasp \emph{the relationship between complexity and information heat capacity} in the context of linguistics, no matter how unconventional this connection may seem.
%%%%%%%%%%%%%%%%%%%%%%%%%%%%%%%%%%%%%%%%%%%%
\subsection{Temperature is a parameter of text complexity}
%%%%%%%%%%%%%%%%%%%%%%%%%%%%%%%%%%%%%%%%%%%
In the previous section, we presented the simple examples of the application of the concept of information temperature. We demonstrated that the entropy and information complexity can both be considered as temperature-dependent functions. Consequently, the temperature itself can be treated as a measure of complexity of the system. We believe that the importance of the concept of information temperature is not limited to these examples and has a much wider scope.

We suppose that the information temperature can be used to characterize the information and intellectual complexity of people and artificial intelligence systems, assuming that they can be represented by written or generated texts. For example, D. Tononi ~\cite{Tononi} indicates the possibility of characterizing a neural network by some function $\Phi$.
This function is not clearly defined and difficult for calculations. Can a neural network be characterized by the temperature of the text that the neural network is able to generate? We suppose that the answer to this question is positive.

%%%%%%%%%%%%%%%%%%%%%%%%%%%%%%%%%%%%%%%%%%%%
\section{Conclusion}\label{Conclusion}
%%%%%%%%%%%%%%%%%%%%%%%%%%%%%%%%%%%%%%%%%%%
In the paper, 
we propose using the information temperature as a measure of complexity and intelligence for an agent capable of generating meaningful texts. The proposed here definition of complexity, which we have called the specific information heat capacity, is expressed in terms of the derivative of entropy with respect to temperature. This specific information heat has a maximum at a finite temperature and, thus, a random sequence is most complex when its information thermocapacity is close to its maximum.
Although the notion of information temperature for random sequences is more complex than the thermodynamic temperature, it is relevant to consider its applicability to natural language and DNA texts. 

We have considered just a special class of binary isotropic  random sequences with a stepwise memory function, 
and the work on the correct introduction of information temperature for random sequences is ongoing. As we know,  the concept of temperature not always can be introduced in thermodynamics as well, and even when it is possible, the system is not always characterized by a single temperature~ \cite{Puglisi}. Similarly, for stationary random sequences, it is important to consider the conditions that permit the introduction of the concept of information temperature.

The next step in the study and development of the concept of information temperature as a measure of complexity is its application to the analysis of symbolic random sequences with arbitrary finite state space and different memory functions. Such sequences cannot be treated in general case as reversible and require the use of methods of modern stochastic thermodynamics for their examination, which is out of the scope of the present paper.

\textbf{Acknowledgments.}
We thank I. Gornyi, Yu. Gefen, and S.S. Melnik for their interest in this study and helpful discussions. Initially, this work was supposed to be done as part of the project of the National Research Foundation of Ukraine, project No. 2021.01/0016.
``Mutual long-range correlations, memory functions, entropy and information temperature of nucleotide sequences of RNA viruses as indicators of danger to humans''.
However, the war in Ukraine led to the suspension of funding for this Project.

%%%%%%%%%%%%%%%%%%%%%%%%%%%%%%%%%%%%%%%%%%%%%%%%%%%%%%%%%%%%%%%%%

\end{document}